\documentclass[10pt]{article}

\usepackage[dvips]{graphicx}
\usepackage{epic}
\usepackage{curves}
\usepackage{amsfonts}

\newcommand{\bela}[1]{\begin{equation}\label{#1}}
\newcommand{\ela}{\end{equation}}
\newcommand{\bear}[1]{\begin{array}{#1}}
\newcommand{\ear}{\end{array}}

\renewcommand{\Psi}{\mbox{\boldmath $\psi$}}

\newcommand{\as}{\\[.6em]}
\newcommand{\As}{\\[.9em]}
\newcommand{\AS}{\\[1.2em]}

\newcommand{\dis}{\displaystyle}

\newcommand{\tr}{\,\mbox{tr}\,}

\newcommand{\del}{\partial}

\newcommand{\eins}{\mbox{\rm 1\hspace{-2.2pt}l}}

\newcommand{\tra}{^{\scriptscriptstyle \mathsf{T}}}

\newcommand{\up}[1]{\vec{#1}}
\newcommand{\down}[1]{\underline{#1}}
\newcommand{\both}[1]{\vec{\underline{#1}}}

\newtheorem{theorem}{Theorem}

\newtheorem{definition}{Definition}

\begin{document}
\begin{center}
  \Large\bf
  On the unification of classical and novel integrable surfaces:\\
  I.\ Differential geometry\\[8mm]
  \large\sc By W.K. Schief and
  B.G.\ Konopelchenko\footnote{Permanent address: Dipartimento di Fisica,
  Universit\`a di Lecce and Sezione INFN, 73100 Lecce, Italy}\\[2mm]
  \small\sl School of Mathematics, The University of New South Wales,\\
  Sydney, NSW 2052, Australia\\[9mm]
\end{center}

\begin{abstract}
A novel class of integrable surfaces is recorded. This class of O surfaces is
shown to include and generalize classical surfaces such as isothermic, constant
mean curvature, minimal, `linear' Weingarten, Guichard and Petot surfaces and 
surfaces of 
constant Gau{\ss}ian curvature. It is demonstrated that the construction of a 
B\"acklund transformation for O~surfaces leads in a natural manner to an
associated parameter-dependent linear representation. The classical
pseudosphere and breather pseudospherical surfaces are generated.
\end{abstract}

\section{Introduction}

It was Jonas (1915) who pointed out that the B\"acklund
transformations and associated permutability theorems which had been
established for a variety of classes of surfaces around the turn of the
century have a common origin. Amongst these transformations are those
by Ribaucour and Koenigs which preserve lines of curvature and
conjugate nets governed by Laplace equations with equal point invariants
respectively. Both transformations
constitute particular cases of a transformation which was termed 
`Fundamental transformation' by Eisenhart (1962) in his treatise 
`Transformations of
Surfaces'. The Fundamental transformation and its cousins
have since then been recognized as central in the geometric and 
algebraic construction of B\"acklund and Darboux-type transformations and their
applications in soliton 
\mbox{theory}~(Matveev \& Salle 1991; Rogers \& Shadwick 1982; Rogers \& Schief
2000). In this connection, 
it is observed that the celebrated coherent structure solutions (dromions) of 
the Davey-Stewartson~I equation (Boiti {\sl et al.\ }1988)
have been derived by a 
Darboux-type transformation which represents nothing but a variant of the
Fundamental transformation.

In an attempt to complement Jonas' fundamental contribution, we here embark on
a study of the common origin of classes of surfaces which are 
invariant under the Fundamental transformation. Thus, we consider sets of $n$ 
surfaces in a Euclidean space $\mathbb{R}^3$ which
are related by the classical Combescure transformation defining parallel 
conjugate nets. These may be 
canonically associated with three Combescure-related surfaces in a
dual \mbox{(pseudo-)Euclidean}
space $\mathbb{R}^n$. We then isolate a privileged 
class of surfaces (O~surfaces) by demanding that the surfaces in both 
$\mathbb{R}^3$ and $\mathbb{R}^n$ be parametrized in terms of orthogonal
coordinates. Remarkably, it turns out that this class of O~surfaces
encapsulates as canonical reductions classical surfaces such as 
isothermic, constant mean curvature, minimal, `linear' Weingarten, 
Guichard and Petot surfaces and surfaces of constant Gau{\ss}ian curvature. 
These are obtained by specifying appropriately the 
dimension and the metric of the dual space~$\mathbb{R}^n$.

It is no accident that Guichard surfaces arise in this context. Thus, 
in a classical note in {\sl Comptes Rendus de l'Acad\'emie des 
Sciences}, Guichard (1900) characterizes these surfaces in the following 
manner:

\medskip
\noindent
{\sl ``Il existe une surface $\rm (N')$ ayant m\^eme image sph\'erique de ses 
lignes de courbure que la surface $\rm (N)$ et telle que si $\rm R_1$ et 
$\rm R_2$ sont les rayons
de courbure principaux de $\rm (N)$, $\rm R_1'$ et $\rm R_2'$ les rayons 
correspondants de $\rm (N')$, on ait 
$$ \rm R_1R'_2 + R_2R_1' = const., $$
la constante n'\'etant pas nulle.''}

\medskip
\noindent
It will be demonstrated that the latter condition may be interpreted as a
particular orthogonality constraint associated with O surfaces. In the
light of this interpretation, Guichard's characterization may be regarded as
containing the essence underlying the definition of O surfaces.

A B\"acklund transformation for O surfaces is obtained by constraining the 
Fundamental transformation in such a way that the above-mentioned 
orthogonality conditions are preserved. As a by-product, a matrix Lax pair for
O~surfaces is derived. 
As an application of the B\"acklund 
transformation for O~surfaces, the classical pseudosphere and breather
pseudospherical surfaces are generated.
The Gau{\ss}-Mainardi-Codazzi equations for O surfaces
are set down and it is shown that these may be regarded as the compatibility
condition for a parameter-dependent linear system which generalizes the 
classical linear representation for isothermic surfaces set down by Darboux 
(1899) (Eisenhart 1962). 

It is important to note that
the formalism developed in this paper may readily be adapted to the case of
integrable difference geometry (Bobenko \& Seiler~1999). This may be regarded
as a first step towards a unified description of 
inte\-grability-preserving discretizations of
differential geometries. In particular,
integrable\linebreak
difference-geometric analogues of the above-mentioned classical 
surfaces are constructed without difficulty. This is the subject of a 
forthcoming article (Schief~2000b).

\section{Conjugate coordinates and the Combescure transformation}

In the following, we are concerned with the geometry of surfaces in
a three-dimensio\-nal Euclidean space. If a surface 
$\Sigma\subset\mathbb{R}^3$ is parametrized in terms of (local) coordinates
$(x,y)$ in such a way that the position vector $\up{R}=\vec{R}(x,y)$ to the
surface~$\Sigma$ obeys a linear hyperbolic equation of the form
\bela{E1}
  \up{R}_{xy} = a\up{R}_x + b\up{R}_y
\end{equation}
then the curves $x=\mbox{\rm const}$ and $y=\mbox{\rm const}$ are 
said to form a {\em conjugate net} (Eisenhart 1962) on~$\Sigma$. In this 
case, it is convenient to introduce the parametrization
\bela{E2}
  a = (\ln H)_y,\quad b = (\ln K)_x
\end{equation}
and tangent vectors $\up{X},\up{Y}$ according to
\bela{E3}
  \up{R}_x = \up{X}H,\quad \up{R}_y = \up{Y}K.
\end{equation}
The second-order equation (\ref{E1}) may then be brought into the first-order
form
\bela{E4}
  \up{X}_y = q\up{Y},\quad \up{Y}_x = p\up{X},
\end{equation}
where the coefficients $p$ and $q$ are defined by
\bela{E5}
  H_y = pK,\quad K_x = qH.
\end{equation}
The latter system may be regarded as {\em adjoint} to the linear system 
(\ref{E4}).

Conversely, if $\{\up{X},\up{Y},H,K\}$ constitutes a solution of the
linear system (\ref{E4}), (\ref{E5}) for some functions $p$ and $q$ then
the relations (\ref{E3}) are compatible and $\up{R}$ may be interpreted as
the position vector of a surface $\Sigma\subset\mathbb{R}^3$ parametrized in 
terms of conjugate coordinates. A second solution $\{H_*,K_*\}$ of the 
adjoint system (\ref{E5}) gives rise to a second surface $\Sigma_*$,
the position vector of which is defined~by
\bela{E6}
  \up{R}_{*x} = \up{X} H_*,\quad \up{R}_{*y} = \up{Y}K_*.
\end{equation}
Accordingly, at corresponding points, the tangent vectors to the 
coordinate lines on the surfaces $\Sigma$ and $\Sigma_*$ are parallel. The 
surface $\Sigma_*$ is termed a {\em Combescure transform} (Eisenhart 1962) 
of the surface $\Sigma$. Hence, the Combescure transformation maps conjugate 
nets to {\em parallel}\footnote{This notion of parallelism is not to be 
confused with the definition of parallel surfaces.} conjugate nets. It is
important to note that the unit normal $\up{N}$ to~$\Sigma$ and its
Combescure transform $\up{N}_{*}$ coincide, that is $\up{N}_{*}=\up{N}$.

It is evident that {\em lines of curvature}
(Eisenhart 1960), which are uniquely defined by the 
requirement that they be conjugate and orthogonal, are also preserved by the 
Combescure transformation. A particular Combescure transform 
$\Sigma_{\circ}$ of a surface~$\Sigma$ parametrized in terms of curvature 
coordinates is therefore
given by its spherical representation, that is the parametrized sphere swept 
out by the unit normal $\up{N}$ to~$\Sigma$. Indeed, the linear system 
(\ref{E4}) implies that
\bela{E6a}
   \up{N}_x\cdot\up{Y} = 0,\quad \up{N}_y\cdot\up{X} = 0
\end{equation}
and hence there exist functions $H_{\circ}$ and $K_{\circ}$ such that
\bela{E6b}
  \up{N}_x = \up{X}H_{\circ},\quad \up{N}_y = \up{Y}K_{\circ}
\end{equation}
by virtue of the orthogonality condition $\up{X}\cdot\up{Y}=0$.
Thus, the coordinate system on~$\Sigma_{\circ}$ generated by the position 
vector $\up{R}_{\circ}=\up{N}$ is conjugate and parallel to that on~$\Sigma$.
In fact, the relations (\ref{E6b}) constitute the well-known Rodrigues
formulae (Eisenhart 1960) if one expresses $H_{\circ}$ and $K_{\circ}$ in
terms of the principal curvatures~(cf.~\S 4).

\section{Combescure-related surfaces and their duals}

It is natural to investigate the properties of sets 
$\{\Sigma_1,\ldots,\Sigma_n$\} of surfaces which are related by Combescure 
transformations. To this end, we consider the linear systems
\bela{E7}
  \bear{rlrl}
    \up{X}_y = & q\up{Y},\quad & \down{H}_y = & p\down{K}\as
    \up{Y}_x = & p\up{X},\quad & \down{K}_x = & q\down{H},
  \ear
\end{equation}
where $\up{X},\up{Y}\in\mathbb{R}^3$ and $\down{H},\down{K}\in\mathbb{R}^n$
are interpreted as column and row vectors respectively, and define a matrix 
$\both{R}\in\mathbb{R}^{3,n}$ via the compatible equations
\bela{E8}
  \both{R}_x = \up{X}\down{H},\quad \both{R}_y = \up{Y}\down{K}.
\end{equation}
Thus, the geometric interpretation given below is immediate:

\medskip
\noindent
{\em The vectors
 $$ \up{R}_{\kappa}\in\mathbb{R}^3,\quad\kappa=1,\ldots,n $$
parametrize parallel conjugate nets on surfaces $\Sigma_{\kappa}\subset
\mathbb{R}^3$ with tangent vectors $\up{X}$ and $\up{Y}$.}

\medskip
\noindent
However, since there exists complete symmetry between $\{\up{X},\up{Y}\}$
and $\{\down{H},\down{K}\}$ and the definition of conjugate nets is in fact
independent of the dimension of the ambient space, the following point of view
is also valid:

\medskip
\noindent
{\em The vectors
 $$ \down{R}^{k}\in\mathbb{R}^n,\quad k=1,2,3 $$
parametrize parallel conjugate nets on surfaces $\Sigma^k\subset
\mathbb{R}^n$ with tangent vectors $\down{H}$ and $\down{K}$.}

\medskip
We refer to the surfaces $\Sigma^k$ as {\em dual} to the surfaces
$\Sigma_{\kappa}$. As mentioned earlier, we here regard the ambient space 
$\mathbb{R}^3$ as a Euclidean space even though
the generalization to pseudo-Euclidean spaces $\mathbb{R}^3$ and
their higher-dimen\-sional analogues is straightforward. 
By contrast, it turns out pivotal to deal with pseudo-Euclidean
dual spaces $\mathbb{R}^n$. Thus, we endow
$\mathbb{R}^n$ with the inner product
\bela{E10}
  \down{H}\cdot\down{K} = \down{H}\,\down{K}\tra = \sum_{\kappa,\mu=1}^n
  H_{\kappa}S^{\kappa\mu}K_{\mu},
\end{equation}
where $S=(S^{\kappa\mu})$ is a constant symmetric matrix. Moreover, it is noted
that the coordinate lines and the associated tangent vectors on the surfaces 
$\Sigma_{\kappa}$ may be reparametrized according to
\bela{E11}
  \bear{rl}
    (\del_x,\up{X},q)\rightarrow & f(x) (\del_x,\up{X},q)\as
    (\del_y,\up{Y},p)\rightarrow & g(y) (\del_y,\up{Y},p)
  \ear
\end{equation}
without changing the tangent vectors $\down{H}$ and $\down{K}$ on the dual
surfaces $\Sigma^k$. Analogously, the change of variables
\bela{E12}
  \bear{rl}
    (\del_x,\down{H},p)\rightarrow & \tilde{f}(x) (\del_x,\down{H},p)\as
    (\del_y,\down{K},q)\rightarrow & \tilde{g}(y) (\del_y,\down{K},q)
  \ear
\end{equation}
preserves the tangent vectors $\up{X}$ and $\up{Y}$.

The concept of dual conjugate nets is implicit in the work of Darboux 
(1910). It has been exploited in the context 
of integrable differential/difference geometries by several authors 
(Konopelchenko \& Schief 1998; Doliwa \& Santini 1999). 
The significance of matrix 
`bilinear potentials' such as $\both{R}$ in connection with the iteration of 
the classical Fundamental transformation and its relatives has also been 
discussed (Schief \& Rogers 1998; Liu \& Ma\~nas 1998).

\section{A novel class of integrable surfaces}

\subsection{The geometry of O surfaces}

Since the Combescure transformation preserves lines of curvature, it is
possible to parametrize simultaneously any set of Combescure-related surfaces 
$\Sigma_{\kappa}\subset\mathbb{R}^3$ in terms of curvature coordinates. 
The orthogonality constraint
\bela{E13}
  \up{X}\cdot\up{Y} = 0
\end{equation}
and the linear system (\ref{E7})$_{1,3}$ then
imply that $\up{X}\cdot\up{X}_y=0$ and
$\up{Y}\cdot\up{Y}_x=0$. An appropriate reparametrization of the
form (\ref{E11}) therefore results in 
\bela{E14}
  \up{X}^2 = 1,\quad \up{Y}^2 = 1.
\end{equation}
Hence, it may be assumed without loss of generality that $\up{X}$ and $\up{Y}$
constitute orthogonal unit vectors. However, the coordinate lines on
the associated dual surfaces $\Sigma^k\subset\mathbb{R}^n$ are not
necessarily orthogonal. In fact, the orthogonality condition
\bela{E15}
  \down{H}\cdot\down{K} = 0
\end{equation}
imposes severe constraints on the surfaces $\Sigma_{\kappa}$. Thus, in this 
manner, one isolates a privileged class of surfaces in $\mathbb{R}^3$. It 
is this class of surfaces that will be the subject of the remainder of the 
present paper.

\begin{definition} {\bf (O surfaces)}\label{def}
Combescure-related parametrized surfaces 
\mbox{$\Sigma_{\kappa}\subset\mathbb{R}^3$} 
and their duals $\Sigma^k\subset\mathbb{R}^n$ are termed {\em (dual) 
O surfaces} if the coordinates on both $\Sigma_{\kappa}$ and 
$\Sigma^k$ are orthogonal. 
\end{definition}

The above terminology is  borrowed from Eisenhart 
(1962) who defines \mbox{\em O nets} as orthogonal conjugate nets.
The adjoint system (\ref{E7})$_{2,4}$ for dual O surfaces implies 
that $\down{H}_y\cdot\down{H}=0$
and $\down{K}_x\cdot\down{K}=0$. Once again, a suitable reparametrization
of the form (\ref{E12}) yields
\bela{E16}
   \down{H}^2 = \pm 1,0,\quad \down{K}^2 = \pm 1,0
\end{equation}
so that the assumption that $\down{H}$ and $\down{K}$ constitute orthogonal
unit or null vectors is admissible.

Before we establish the integrability of O surfaces in the sense of the 
existence of a parameter-dependent linear representation and an associated
B\"acklund transformation, we demonstrate below how classical surfaces such
as isothermic, constant mean curvature, minimal, `linear' Weingarten, Guichard
and Petot 
surfaces and surfaces of constant Gau{\ss}ian curvature may be retrieved as 
canonical examples of O~surfaces.

\subsection{Examples} 

In view of the following, it is recalled that
if a surface $\Sigma\subset\mathbb{R}^3$ is parametrized in terms 
of curvature coordinates then the associated
{\em principal curvatures} (Eisenhart 1960) $\mathsf{h}$~and~$\mathsf{k}$ 
read 
\bela{E17}
 \bear{rcl}
 \mathsf{h} = &\dis
      -\frac{\up{R}_{x}\cdot\up{N}_{x}}{
        \up{R}_{x}^2} = & \dis -\frac{H_{\circ}}{H}\AS
 \mathsf{k} = &\dis
      -\frac{\up{R}_{y}\cdot\up{N}_{y}}{
        \up{R}_{y}^2} = & \dis -\frac{K_{\circ}}{K},
 \ear
\end{equation}
where $H_{\circ}$ and $K_{\circ}$ are defined by (\ref{E6b}).
In particular, the {\em Gau{\ss}ian} and {\em mean curvatures} $\mathsf{K}$ and 
$\mathsf{M}$ respectively of the surface $\Sigma$ are given by
\bela{E18}
 \bear{rl}
   \mathsf{K} = &\mathsf{h}\mathsf{k} = 
   \dis \frac{H_{\circ}K_{\circ}}{H K}\AS
   \mathsf{M} = &\mathsf{h}+\mathsf{k} = 
   \dis -\frac{H_{\circ}}{H} - \frac{K_{\circ}}{K}.
 \ear  
\end{equation}

\subsubsection{Surfaces of constant Gau{\ss}ian curvature} 

We first consider the simplest choice
\bela{E19} 
   S = \left(\bear{cc}1&0\\ 0&1\ear\right)
\end{equation}
which corresponds to a two-dimensional Eulidean dual space $\mathbb{R}^2$.
In this case, the orthogonality condition (\ref{E15}) takes the form
\bela{E20}
  H_1K_1 + H_2K_2 = 0.
\end{equation}
By virtue of (\ref{E18})$_1$, this is equivalent to the requirement that the
Gau{\ss}ian curvatures of $\Sigma_1$ and $\Sigma_2$ be related by
\bela{E21}
  \mathsf{K}_1 = -\mathsf{K}_2.
\end{equation}
Alternatively, if we consider a pseudo-Euclidean dual space $\mathbb{R}^2$
with 
\bela{E22}
   S = \left(\bear{cc}1&0\\ 0&-1\ear\right)
\end{equation}
then the orthogonality condition reads 
\bela{E23}
  H_1K_1 = H_2K_2
\end{equation}
so that
\bela{E24}
  \mathsf{K}_1 = \mathsf{K}_2.
\end{equation}
We therefore conclude that a pair of Combescure-related surfaces parametrized
in terms of curvature coordinates constitute O surfaces if, at corresponding 
points, their Gau{\ss}ian curvatures are of the same magnitude. In particular,
if we
restrict the surface $\Sigma_2$ to the sphere with $\mathsf{K}_2=1$ then the 
surface $\Sigma_1$ is of constant Gau{\ss}ian curvature and $\Sigma_2$ is but 
its spherical representation. Thus, classical {\em (pseudo)spherical 
surfaces}~(Eisenhart 1960) are retrieved.

\subsubsection{Isothermic and minimal surfaces} 

The choice
\bela{E25}
  S=\left(\bear{cc}0&1\\ 1&0\ear\right)
\end{equation}
leads to the orthogonality condition
\bela{E26}
  H_1K_2 + H_2K_1 = 0.
\end{equation}
The adjoint system (\ref{E7})$_{2,4}$ then implies that $(H_1H_2)_y=0$ and
$(K_1K_2)_x=0$. Without loss of generality, we may set
\bela{E28}
  H_1H_2=1,\quad K_1K_2=-1
\end{equation}
so that
\bela{E29}
  K_1 = H_1,\quad K_2 = -H_2.
\end{equation}
In terms of the position vectors $\up{R}_1$ and $\up{R}_2$ to the
surfaces $\Sigma_1$ and $\Sigma_2$, the relations (\ref{E28}), (\ref{E29})
translate into
\bela{E30}
  \up{R}_{2x} = \frac{\up{R}_{1x}}{\up{R}_{1x}^2},\quad
  \up{R}_{2y} = - \frac{\up{R}_{1y}}{\up{R}_{1y}^2}
\end{equation}
and the {\em conformality} condition
\bela{E31}
  \up{R}_{1x}^2 = \up{R}_{1y}^2,\quad \up{R}_{2x}^2 = \up{R}_{2y}^2.
\end{equation}
Thus, the surfaces $\Sigma_1$ and $\Sigma_2$ constitute classical
{\em isothermic surfaces} (Eisenhart 1960) which are related by the
{\em Christoffel transformation} (Eisenhart 1962) (\ref{E30}). Furthermore,
if we identify~$\Sigma_2$ with the spherical representation of $\Sigma_1$ then
$H_2=H_{\circ}$ and $K_2=K_{\circ}$ which, in turn, implies that the 
surface $\Sigma_1$ is {\em minimal} (Eisenhart 1960) since
\bela{E32}
  \mathsf{M}_1=-\frac{H_2}{H_1} - \frac{K_2}{K_1} = 0.
\end{equation}
The latter is in agreement with the well-known fact that the Christoffel
transform of a minimal surface constitutes a sphere. It is also noted that
isothermic surfaces in spaces of arbitrary dimension (Schief 2000a) may be 
retrieved by considering O~surfaces in~$\mathbb{R}^m$.

\subsubsection{Constant mean curvature surfaces and the Bonnet theorem} 

In the 
preceding, we have shown how classical minimal surfaces are obtained within
the framework of O surfaces. The important class of {\em constant mean
curvature surfaces} (Eisenhart 1960), 
which constitute particular isothermic surfaces, may also
be retrieved in a natural manner. Thus, we observe that any set of 
Combescure-related O surfaces $\Sigma_{\kappa}$ gives rise to an infinite
number of Combescure-related O~surfaces by taking linear combinations of the
associated position vectors $\up{R}_{\kappa}$. For instance, if $\Sigma_1$
and $\Sigma_2$ are two isothermic surfaces related by the Christoffel
transformation then the surfaces $\Sigma_{\pm}$ with position vectors
\bela{E33}
  \up{R}_{\pm} = \frac{1}{2}(\up{R}_2\pm\up{R}_1)
\end{equation}
constitute O surfaces which are Combescure transforms of both $\Sigma_1$ and  
$\Sigma_2$. The corresponding solutions 
of the adjoint system (\ref{E7})$_{2,4}$ are given by
\bela{E34}
   H_{\pm} = \frac{1}{2}(H_2\pm H_1),\quad K_{\pm} = \frac{1}{2}(K_2\pm K_1).
\end{equation}
Accordingly, the Gau{\ss}ian curvatures of the surfaces $\Sigma_{\pm}$
take the form
\bela{E35}
  \mathsf{K}_{\pm} = \frac{H_{\circ}K_{\circ}}{H_{\pm}K_{\pm}}
           = \frac{4 H_{\circ}K_{\circ}}{H_1K_1 + H_2K_2}
\end{equation}
by virtue of (\ref{E26}) and hence coincide. 
This is not surprising since the transition from  
$(\Sigma_1,\Sigma_2)$ to $(\Sigma_+,\Sigma_-)$ may be interpreted at the
level of the matrix $S$ as a similarity transformation  mapping the case
(\ref{E25}) to the case (\ref{E22}).

If we now identify the surface $\Sigma_-$ with the spherical representation
of the isothermic surfaces, that is
\bela{E36}
   \up{R}_- = \up{N},\quad H_- = H_{\circ},\quad K_- = K_{\circ},
\end{equation}
then
\bela{E37}
  \mathsf{K}_{\pm}=1,\quad \mathsf{M}_1=1,\quad \mathsf{M}_2=-1.
\end{equation}
The latter relations encapsulate a well-known theorem due to Bonnet
(Eisenhart 1960) which states that with any surface $\Sigma_+$ of constant 
Gau{\ss}ian curvature $\mathsf{K}_+=1$ one may associate two parallel
surfaces $\Sigma_1$ and $\Sigma_2$ of constant mean curvature 
$\mathsf{M}_1=1$ and~$\mathsf{M}_2=-1$ respectively
with position vectors
\bela{E38}
  \up{R}_1 = \up{R}_+ - \up{N},\quad \up{R}_2 = \up{R}_+ + \up{N}.
\end{equation}
This may be regarded as a special case of the following statement:

\medskip\noindent
{\em If the Gau{\ss}ian curvatures of two Combescure-related surfaces 
$\Sigma_{\pm}$ parametrized in terms of curvature coordinates are equal at 
corresponding points then
the Combescure transforms $\Sigma_1$ and $\Sigma_2$ defined by
\bela{E39}
  \up{R}_1 = \up{R}_+ - \up{R}_-,\quad \up{R}_2 = \up{R}_+ + \up{R}_-
\end{equation}
constitute isothermic surfaces which are related by the Christoffel
transformation.}

\subsubsection{`Linear' Weingarten surfaces} 

Surfaces of constant Gau{\ss}ian
or mean curvature represent particular examples of 
{\em Weingarten surfaces} (Eisenhart 1960), that is surfaces in 
$\mathbb{R}^3$ which admit
a functional relation between the principal curvatures. `Linear' Weingarten 
surfaces are those corresponding to a functional relation of the form
\bela{39a}
  \alpha\mathsf{K} + \beta\mathsf{M} = \gamma,
\end{equation}
where $\alpha,\beta$ and $\gamma$ are arbitrary constants. If $\Sigma$ 
constitutes a linear Weingarten surface parametrized in terms of 
curvature coordinates and $\Sigma_{\circ}$ denotes its spherical representation
then, on use of the expressions (\ref{E18}) for the Gau{\ss}ian and mean 
curvatures $\mathsf{K}$ and $\mathsf{M}$, the above relation may be brought
into the form
\bela{E39b}
  \down{H}\cdot\down{K} = 0,\quad S = \left(\bear{cc}\gamma & \beta\\
  \beta & -\alpha\ear\right)
\end{equation}
with the identification
\bela{E39c}
  \down{H} = (H,H_{\circ}),\quad\down{K} = (K,K_{\circ}).
\end{equation}
Thus, linear Weingarten surfaces constitute O surfaces which are parallel
to surfaces of constant Gau{\ss}ian curvature since the matrix $S$ as given by
(\ref{E39b})$_2$ may be mapped by means of an appropriate similarity 
transformation to either (\ref{E19}) or~(\ref{E22}) provided that 
$\det S\neq 0$. At the level of the position matrix $\both{R}$, this
corresponds to a linear transformation.
   
\subsubsection{Guichard surfaces} 

If we equip a three-dimensional dual space with
the indefinite metric
\bela{E40}
  S = \left(\bear{ccc}0&1&0\\ 1&0&0\\ 0&0&1\ear\right)
\end{equation}
then the orthogonality condition becomes
\bela{E41}
  H_1K_2 + H_2K_1 + H_3K_3 = 0.
\end{equation}
The corresponding O surfaces in $\mathbb{R}^3$ 
evidently generalize isothermic surfaces. If 
$\Sigma_3$ is taken to be the spherical representation of $\Sigma_1$ and 
$\Sigma_2$ then the relations $H_3=H_{\circ}$ and $K_3=K_{\circ}$ imply that
\bela{E42}
  \frac{1}{\mathsf{h}_1\mathsf{k}_2} + 
  \frac{1}{\mathsf{h}_2\mathsf{k}_1} + 1 = 0.
\end{equation} 
Accordingly, $\Sigma_1$ and $\Sigma_2$ represent {\em Guichard surfaces}
(Guichard 1900; Eisenhart 1962) as alluded to in the introduction.

\subsubsection{Petot surfaces} 

Another canonical class of O surfaces is
obtained by identifying the three-dimensional Euclidean space with its dual.
Thus, if we set
\bela{E43}
  \down{H} = \up{X}\tra,\quad \down{K} = \up{Y}\tra,\quad p=q
\end{equation}
then the linear systems (\ref{E7}) coincide. The constraint (\ref{E43})$_3$ is
known to define {\em Petot surfaces} (Petot 1891). 
Accordingly,
the O surfaces $\Sigma_{\kappa}$ constitute three Petot surfaces 
which are linked by Combescure transformations. Moreover, the
defining relations for the `position matrix' $\both{R}$ read
\bela{E44}
  \both{R}_x = \up{X}\up{X}\tra,\quad \both{R}_y = \up{Y}\up{Y}\tra
\end{equation}
which, in turn, imply the `conservation laws'
\bela{E45}
  {(\up{R}_{\kappa x}^2)}_y = {(\up{R}_{\kappa y}^2)}_x.
\end{equation}
The metrics on the surfaces $\Sigma_{\kappa}$ may therefore be derived
from potentials, that is
\bela{E46}
   ds_{\kappa}^2 = d\up{R}_{\kappa} \cdot d\up{R}_{\kappa} = 
      \varphi_{\kappa x}dx^2 + \varphi_{\kappa y}dy^2.
\end{equation}
The latter property constitutes an alternative characterization of
Petot surfaces. In fact, it reflects the fact that Petot surfaces
represent the constituent members of Darboux-Egorov-type
triply orthogonal systems of surfaces (Egorov 1900, 1901).
It is noted, however, that the particular form of the 
potentials $\varphi_{\kappa}$, namely\footnote{For simplicity, we here
use the normalization (\ref{E14}).}
\bela{E47}
  \varphi_{\kappa} = R_{\kappa}^{\kappa},
\end{equation}
indicates that the above Petot surfaces are not generic. 
The class of Petot O surfaces nevertheless enshrines 
the generic class of Petot surfaces in the sense that any Petot
surface may be obtained from a Petot O surface by application of an
appropriate Combescure transformation.

\section{A B\"acklund transformation for O surfaces}

An extensive account of the transformation theory of conjugate nets is 
contained in the treatise `Transformations of Surfaces' by Eisenhart 
(1962). Here, we focus on the classical 
{\em Fundamental transformation} 
(Jonas 1915, Eisenhart 1962). Since 
the Fundamental transformation
commutes with the Combescure transformation, it can be simultaneously applied
to sets of Combescure-related surfaces. 

\subsection{The Fundamental and Ribaucour transformations}

The Fundamental transformation
is generated by two pairs of scalar solutions of the linear 
systems~(\ref{E7}) and 
corresponding bilinear potentials of the form (\ref{E8}). Thus, for a 
given pair of functions $p,q$ associated with a set of Combescure-related 
surfaces $\Sigma_{\kappa}$, let $\{X,Y\}$ and $\{H,K\}$ be solutions
of the linear systems
\bela{E48}
  \bear{rlrl}
    X_y = & qY,\quad & H_y = & pK\as
    Y_x = & pX,\quad & K_x = & qH.
  \ear
\end{equation}
In the sequel, we refer to $X,Y$ and $H,K$ as {\em eigenfunctions} and {\em 
adjoint eigenfunctions} respectively. Three bilinear potentials 
$\up{M},\down{M}$ and $M$ may now be introduced according to
\bela{E49}
 \bear{rlrlrl}
   \up{M}_x = & \up{X}H,\quad & \down{M}_x = & X\down{H},\quad & M_x = & XH\as
   \up{M}_y = & \up{Y}K,\quad & \down{M}_y = & Y\down{K},\quad & M_y = & YK.
 \ear
\end{equation}
A second set of Combescure-related surfaces $\Sigma_{\kappa}'$ is now 
obtained as follows:

\begin{theorem}\label{fundamental}
{\bf (The Fundamental transformation)} The linear systems (\ref{E7}) and the
defining relations (\ref{E8}) are invariant under
\bela{E50}
  (\both{R},\up{X},\up{Y},\down{H},\down{K},p,q)\rightarrow
  (\both{R}',\up{X}',\up{Y}',\down{H}',\down{K}',p',q'),
\end{equation}
where
\bela{E51}
  \both{R}' = \both{R} - \frac{\up{M}\down{M}}{M}
\end{equation}
and
\bela{E52}
 \bear{rlrl}
   \up{X}' = & \dis \up{X} - \frac{X\up{M}}{M},\quad & 
   \up{Y}' = & \dis \up{Y} - \frac{Y\up{M}}{M}\AS
   \down{H}' = & \dis \down{H} - \frac{H\down{M}}{M},\quad &
   \down{K}' = & \dis \down{K} - \frac{K\down{M}}{M}\AS
   p' = & p - \dis \frac{YH}{M},\quad & q' = & q - \dis \frac{XK}{M}.
 \ear
\end{equation}
\end{theorem}
\noindent  
It is emphasized that the above transformation may also be regarded as a 
mapping between the sets of dual surfaces $\Sigma^k$ and $\Sigma^{\prime k}$. 

If the coordinate lines on $\Sigma_{\kappa}$ are lines of curvature, that
is $\up{X}\cdot\up{Y}=0$, then it is readily verified that the quantities
\bela{E53}
  X = \up{M}\cdot\up{X},\quad Y = \up{M}\cdot\up{Y}
\end{equation}
constitute particular eigenfunctions. This choice of eigenfunctions in the
definitions of the bilinear potentials $\up{M}$ and $M$ leads, in turn, to
the relations
\bela{E54}
  {(\up{M}^2)}_x = 2M_x,\quad {(\up{M}^2)}_y = 2M_y
\end{equation}
so that we may set
\bela{E55}
  \up{M}^2 = 2M.
\end{equation}
It is now straightforward to show that
\bela{E56}
  \up{X}^{\prime2} = \up{X}^2,\quad \up{X}'\cdot\up{Y}' = 0,\quad
  \up{Y}^{\prime2} = \up{Y}^2.
\end{equation}
Thus, it turns out that lines of curvature and the normalisation
(\ref{E14}) are preserved by the Fundamental transformation if the
eigenfunctions $X,Y$ and the bilinear potential~$M$ are chosen to be
(\ref{E53}) and (\ref{E55}) respectively. Under these circumstances, the
Fundamental transformation becomes the classical {\em Ribaucour 
transformation}~(Eisenhart 1962).

\subsection{Application to O surfaces}

It is remarkable that the Ribaucour transformation may be constrained in such
a way that orthogonality of the coordinate lines on the dual surfaces  
is also sustained. In fact, as a by-product, a parameter-dependent linear 
representation of O surfaces is obtained. As in the preceding, we first observe
that the quantities
\bela{E57}
  H = \lambda\down{M}\cdot\down{H},\quad K = \lambda\down{M}\cdot\down{K}
\end{equation}
constitute particular adjoint eigenfunctions. The constant parameter 
$\lambda$ is
now non-trivial as we have already specified the eigenfunctions $X$ and $Y$.
The associated potentials $\down{M}$ and $M$ then obey the relations
\bela{E58}
  \lambda{(\down{M}^2)}_x = 2M_x,\quad
  \lambda{(\down{M}^2)}_y = 2M_y
\end{equation}
so that
\bela{E59}
  \lambda\down{M}^2 = 2M
\end{equation}
is, at least, consistent. It is shown below that this constraint is indeed
admissible. On this assumption, we now proceed and note that
\bela{E60}
  \down{H}^{\prime2} = \down{H}^2,\quad \down{H}'\cdot\down{K}' = 0,\quad
  \down{K}^{\prime2} = \down{K}^2,
\end{equation}
which implies that the coordinate lines are also orthogonal on the
transformed dual surfaces $\Sigma^{\prime k}$ 
with the normalisation (\ref{E16}) unchanged.
 
Finally, insertion of the (adjoint) eigenfunctions $X,Y$ and $H,K$ as given
by (\ref{E53}) and (\ref{E57}) respectively into the defining relations
(\ref{E49}) produces the following {\em Lax pair} for O surfaces:

\begin{theorem}\label{lax}
{\bf (A Lax pair for O surfaces)} The linear system
\bela{E61}
 \bear{rl}
  \left(\bear{c}\up{M}\\ \down{M}\tra\ear\right)_x = &
  \left(\bear{cc}0&\lambda\up{X}\down{H}\\ \down{H}\tra\up{X}\tra&0\ear\right)
  \left(\bear{c}\up{M}\\ \down{M}\tra\ear\right)
  \\[5mm]
  \left(\bear{c}\up{M}\\ \down{M}\tra\ear\right)_y = &
  \left(\bear{cc}0&\lambda\up{Y}\down{K}\\ \down{K}\tra\up{Y}\tra&0\ear\right)
  \left(\bear{c}\up{M}\\ \down{M}\tra\ear\right)
 \ear
\end{equation}
is compatible modulo the linear systems (\ref{E7}) and the orthogonality
conditions (\ref{E13}) and (\ref{E15}). It admits the first integral
\bela{E62}
  \up{M}^2 - \lambda\down{M}^2 = \mbox{\rm const}.
\end{equation}
\end{theorem}

\noindent
The existence of the first integral (\ref{E62}) guarantees that the constraint 
(\ref{E59}) is admissible. Consequently, we are now in a position to formulate 
the following theorem:

\begin{theorem}\label{back}
{\bf (A B\"acklund transformation for O surfaces)} Let $\both{R}$ be the
position matrix of a set of Combescure-related O surfaces $\Sigma_{\kappa}$ and
their duals $\Sigma^k$ and $\up{X},\up{Y},\down{H},\down{K}$ corresponding
tangent vectors. If the vectors $\up{M}$ and $\down{M}$ constitute a solution
of the linear system (\ref{E61}) subject to the admissible constraint
\bela{E63}
  \up{M}^2 = \lambda\down{M}^2 = 2M
\end{equation}
and the scalar $M$ is defined by the latter then the position matrix of a 
second set of O surfaces $\Sigma'_{\kappa},\Sigma^{\prime k}$ is given by
\bela{E64}
  \both{R}' = \both{R} - \frac{\up{M}\down{M}}{M}.
\end{equation}
\end{theorem}

We conclude this section with two remarks. Firstly, if the O surface $\Sigma_n$
is identified with the spherical representation of the remaining O surfaces
$\Sigma_{\kappa}$ then
\bela{E65}
  \up{R}_n^2 = 1.
\end{equation}
In this case, by differentiation, it is readily verified that
\bela{E66}
  M_n = \up{R}_n\cdot\up{M}
\end{equation}
is another admissible constraint. Consequently, the $n$th component of the
transformation law (\ref{E64}) may be cast into the form
\bela{E67}
  \up{R}'_n = \left(\eins - 2\frac{\up{M}\up{M}\tra}{\up{M}^2}\right)\up{R}_n
\end{equation}
which implies that
\bela{E68}
  \up{R}^{\prime2}_n = 1.
\end{equation}
Hence, we come to the important conclusion that the above B\"acklund 
transformation acts within specific sub-classes of O surfaces such
as (pseudo)spherical, minimal or Guichard surfaces. Moreover, it is readily
shown that constraints of the form
\bela{E78a}
  \left(\sum_{\kappa=1}^n c_{\kappa}\up{R}_{\kappa}\right)^2=1,
\end{equation}
which generalize (\ref{E65}), may also be preserved. In particular, the
specialization (\ref{E36}) leading to constant mean curvature 
surfaces proves invariant.

Secondly, in Hertrich-Jeromin \& Pedit (1997), it has been pointed out that
the classical B\"acklund transformation for isothermic sufaces may be 
formulated in terms of a {\em matrix Riccati system}. It turns out that
such a system is generic. Thus, differentiation
of (\ref{E64}) and use of (\ref{E8}) lead to a matrix Riccati system
for the new position matrix~$\both{R}'$, namely
\bela{E69}
  {(\Delta\both{R})}_{x^i} = -\frac{\lambda}{2}(\both{R}_{x^i}
   \Delta\both{R}\tra
  \Delta\both{R} + \Delta\both{R}\Delta\both{R}\tra\both{R}_{x^i})
  + \lambda\Delta\both{R}\tr(\both{R}_{x^i}\Delta\both{R}\tra),
\end{equation}
where $\Delta\both{R}=\both{R}'-\both{R}$ and $(x^1,x^2)=(x,y)$. It is 
emphasized that, remarkably, the above first-order system only involves 
the seed position matrix, its B\"acklund transform and an
arbitrary constant B\"acklund parameter. If one sets aside the genesis of the 
pair (\ref{E69}), it would be interesting to 
determine whether its compatibilty conditions imply that
$\both{R}'$ and $\both{R}$ necessarily define O surfaces. In the terminology
of integrable systems, a pair of this type is termed a {\em strong}
B\"acklund transformation.

\section{The Gau{\ss}-Mainardi-Codazzi equations}

In the preceding, contact has been made with classical isothermic surfaces
in the context of particular O surfaces and a Riccati-type
formulation of the B\"acklund transformation for O surfaces.
It turns out that this connection may be exploited to obtain an alternative
linear representation for O surfaces which generalizes that for isothermic
surfaces obtained by Darboux (1899) (Eisenhart 1962). 
We first note that the 
{\em Gau{\ss}-Weingarten equations} (Eisenhart 1960) for surfaces in 
$\mathbb{R}^3$
parametrized in terms of lines of curvature are encoded in the linear system
\bela{E70}
 \bear{rl}
  \left(\bear{c}\up{X}\\ \up{Y}\\ \up{N}\ear\right)_x = &
  \left(\bear{ccc}0 & -p & -H_{\circ}\\p & 0 & 0\\ H_{\circ} & 0 & 0\ear\right)
  \left(\bear{c}\up{X}\\ \up{Y}\\ \up{N}\ear\right)\\[7mm]
  \left(\bear{c}\up{X}\\ \up{Y}\\ \up{N}\ear\right)_y = &
  \left(\bear{ccc}0 & q & 0\\ -q & 0 & -K_{\circ}\\ 0& K_{\circ} & 0\ear\right)
  \left(\bear{c}\up{X}\\ \up{Y}\\ \up{N}\ear\right)
 \ear
\end{equation}
due to the orthogonality condition $\up{X}\cdot\up{Y}=0$. The compatibility 
condition for this system produces the {\em Gau{\ss}-Mainardi-Codazzi
equations}
\bela{E71}
  p_y + q_x + H_{\circ}K_{\circ}=0,\quad H_{\circ y} = pK_{\circ},\quad
  K_{\circ x} = qH_{\circ}.
\end{equation}
If the surfaces constitute Combescure-related O surfaces $\Sigma_{\kappa}$ 
then the underdetermined system (\ref{E71}) is coupled with the relations 
\bela{E72}
  \down{H}\cdot\down{K} = 0,\quad \down{H}_y = p\down{K},\quad
  \down{K}_x = q\down{H}
\end{equation}
which completes the set of Gau{\ss}-Mainardi-Codazzi equations. It is now
directly verified that the system (\ref{E71}), (\ref{E72}) holds 
if and only if the linear system
\bela{E73}
 \bear{rl}
  \left(\bear{c}\mathcal{X}\\ \mathcal{Y}\\ \mathcal{N}\\ 
        \down{\mathcal{R}}\tra\ear\right)_x = &
  \left(\bear{cccc} 0 & -p & -H_{\circ} & m\down{H}\\
                    p & 0  &  0         & 0   \\
                    H_{\circ} & 0 & 0 & 0\\
                    \down{H}\tra & 0 & 0 & 0\ear\right)
  \left(\bear{c}\mathcal{X}\\ \mathcal{Y}\\ \mathcal{N}\\ 
        \down{\mathcal{R}}\tra\ear\right)\\[9mm]
  \left(\bear{c}\mathcal{X}\\ \mathcal{Y}\\ \mathcal{N}\\ 
        \down{\mathcal{R}}\tra\ear\right)_y = &
  \left(\bear{cccc} 0 & q & 0 & 0\\
                    -q & 0  &  -K_{\circ}  & m\down{K}\\
                    0 & K_{\circ} & 0 & 0\\
                    0 &\down{K}\tra & 0 & 0\ear\right)
  \left(\bear{c}\mathcal{X}\\ \mathcal{Y}\\ \mathcal{N}\\ 
        \down{\mathcal{R}}\tra\ear\right)
 \ear
\end{equation}
is compatible. Here, $m$ is an arbitrary constant parameter. Moreover, for
$m=0$, the Gau{\ss}-Weingarten equations (\ref{E70}) together with the
defining relations (\ref{E8}) for the position matrix of O surfaces are 
retrieved if one considers vector-valued solutions of the linear
system (\ref{E73}). It is also noted that the above
parameter-dependent linear representation for O surfaces indeed reduces to
that for isothermic surfaces in the case (\ref{E25}).

As an illustration, we focus on the case (\ref{E19}), that is 
\bela{E74}
  \down{H}^2 = 1,\quad \down{K}^2 = 1,\quad 
  S = \left(\bear{cc}1&0\\ 0&1\ear\right).
\end{equation}
A convenient parametrization of the tangent vectors $\down{H}$ and $\down{K}$
is then given by
\bela{E75}
  H_1 = -K_2 = \cos\theta,\quad K_1 = H_2 = \sin\theta.
\end{equation}
Accordingly, the system (\ref{E72}) reduces to 
\bela{E76}
  p = -\theta_y,\quad q = \theta_x
\end{equation}
and the Gau{\ss}-Mainardi-Codazzi equations become
\bela{E77}
  \theta_{xx} - \theta_{yy} + H_{\circ}K_{\circ} = 0,\quad
  H_{\circ y} = -\theta_yK_{\circ},\quad K_{\circ x} = \theta_x H_{\circ}.
\end{equation}
If we now identify the surface $\Sigma_2$ with the spherical representation of
$\Sigma_1$ then $H_{\circ}=H_2,\,K_{\circ}=K_2$ and the classical sine-Gordon
equation
\bela{E78}
  \theta_{xx} - \theta_{yy} = \sin\theta\cos\theta
\end{equation}
underlying pseudospherical surfaces is recovered (Eisenhart 1960).

\section{The pseudosphere and breather pseudospherical surfaces}

\begin{figure}[h]
 \centerline{\includegraphics[angle=270,trim= 110 190 110 190,
              width=0.49\textwidth,clip]{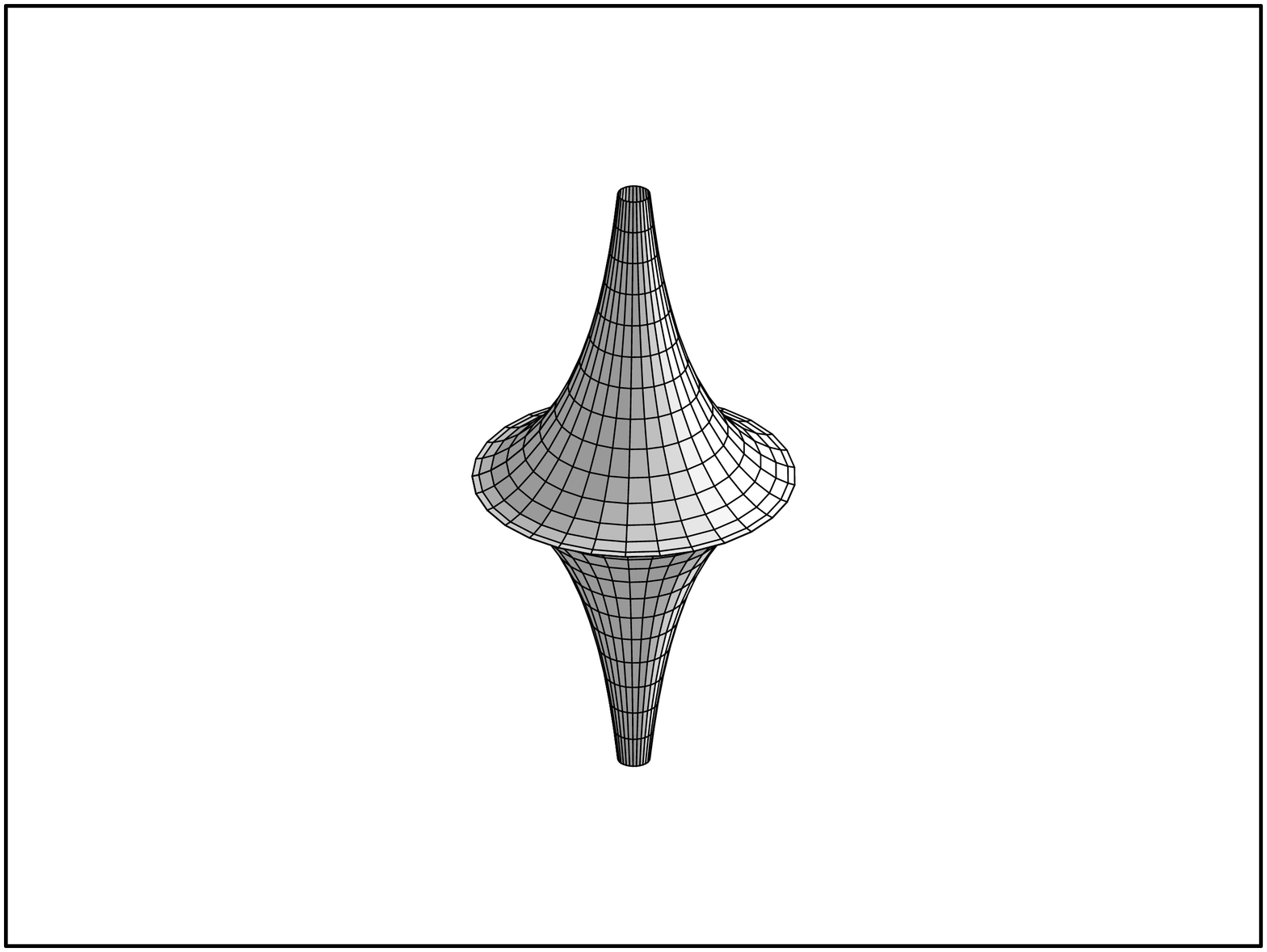}}
 \caption{The classical pseudosphere}
 \label{pseudosphere}
\end{figure}

We conclude this paper with an illustration of the B\"acklund transformation
for O~surfaces and consider the particular case of pseudospherical 
surfaces as discussed in the previous section. Thus, we here regard a straight
line as a (degenerate) seed pseudospherical surface $\Sigma_1$ together with 
its `spherical representation' $\Sigma_2$ represented~by
\bela{E79}
  \up{R}_1 = \left(\bear{c}0\\ 0\\ x\ear\right),\quad
  \up{R}_2 = \left(\bear{c}-\sin y\\ \cos y\\ 0\ear\right)
\end{equation}
so that the tangent vectors to $\Sigma_1,\Sigma_2$ and their duals read
\bela{E80}
  \up{X} = \left(\bear{c}0\\ 0\\ 1\ear\right),\quad
  \up{Y} = \left(\bear{c}\cos y\\ \sin y\\ 0\ear\right),\quad
  \down{H} = (1\,\,\,0),\quad \down{K} = (0\,\,\,-1).
\end{equation}
It is evident that the linear systems (\ref{E7}) with $p=q=0$ and the 
orthogonality conditions $\up{X}\cdot\up{Y}=\down{H}\cdot\down{K}=0$ are
satisfied. Accordingly, the linear system (\ref{E61}) for these particular
O surfaces becomes
\bela{E81}
 \bear{rl}
  \left(\bear{c}M^1\\ M^2\\ M^3\\ M_1\\ M_2\ear\right)_x = &
  \left(\bear{ccccc}0&0&0&0&0\\ 0&0&0&0&0\\ 0&0&0&\lambda&0\\
                    0&0&1&0&0\\ 0&0&0&0&0\ear\right)
  \left(\bear{c}M^1\\ M^2\\ M^3\\ M_1\\ M_2\ear\right)\\[12mm]
  \left(\bear{c}M^1\\ M^2\\ M^3\\ M_1\\ M_2\ear\right)_y = &
  \left(\bear{ccccc}0&0&0&0&-\lambda\cos y\\ 0&0&0&0&-\lambda\sin y\\ 
                    0&0&0&0&0\\ 0&0&0&0&0\\ -\cos y&-\sin y&0&0&0\ear\right)
  \left(\bear{c}M^1\\ M^2\\ M^3\\ M_1\\ M_2\ear\right).
 \ear
\end{equation}
The latter decouples into two systems of linear ordinary differential
equations for $M^3(x),M_1(x)$ and $M^1(y),M^2(y),M_2(y)$ respectively.
The constants of integration in the general solution of (\ref{E81}) have to 
be chosen in such a way that the admissible constraints (\ref{E63}) and 
(\ref{E66})$_{n=2}$ are satisfied. For brevity, we here only state the
result of this analysis:

In the case $\lambda=1$, the position vector $\up{R}'_1$ of the B\"acklund 
transform $\Sigma'_1$ may be reduced to
\bela{E82}
  \up{R}'_1 = \left(\bear{c}\dis\frac{\sin y}{\cosh x}\AS
                         -\dis\frac{\cos y}{\cosh x}\As
                         x - \tanh x\ear\right).
\end{equation}
This pseudsospherical surface of revolution is nothing but
Beltrami's classical {\em pseudosphere} (Eisenhart 1960) as depicted in
Figure \ref{pseudosphere}.

\begin{figure}[h]
 \begin{minipage}{0.49\textwidth}
 \centerline{\includegraphics[angle=270,trim= 110 190 110 190,width=\textwidth,
                              clip]{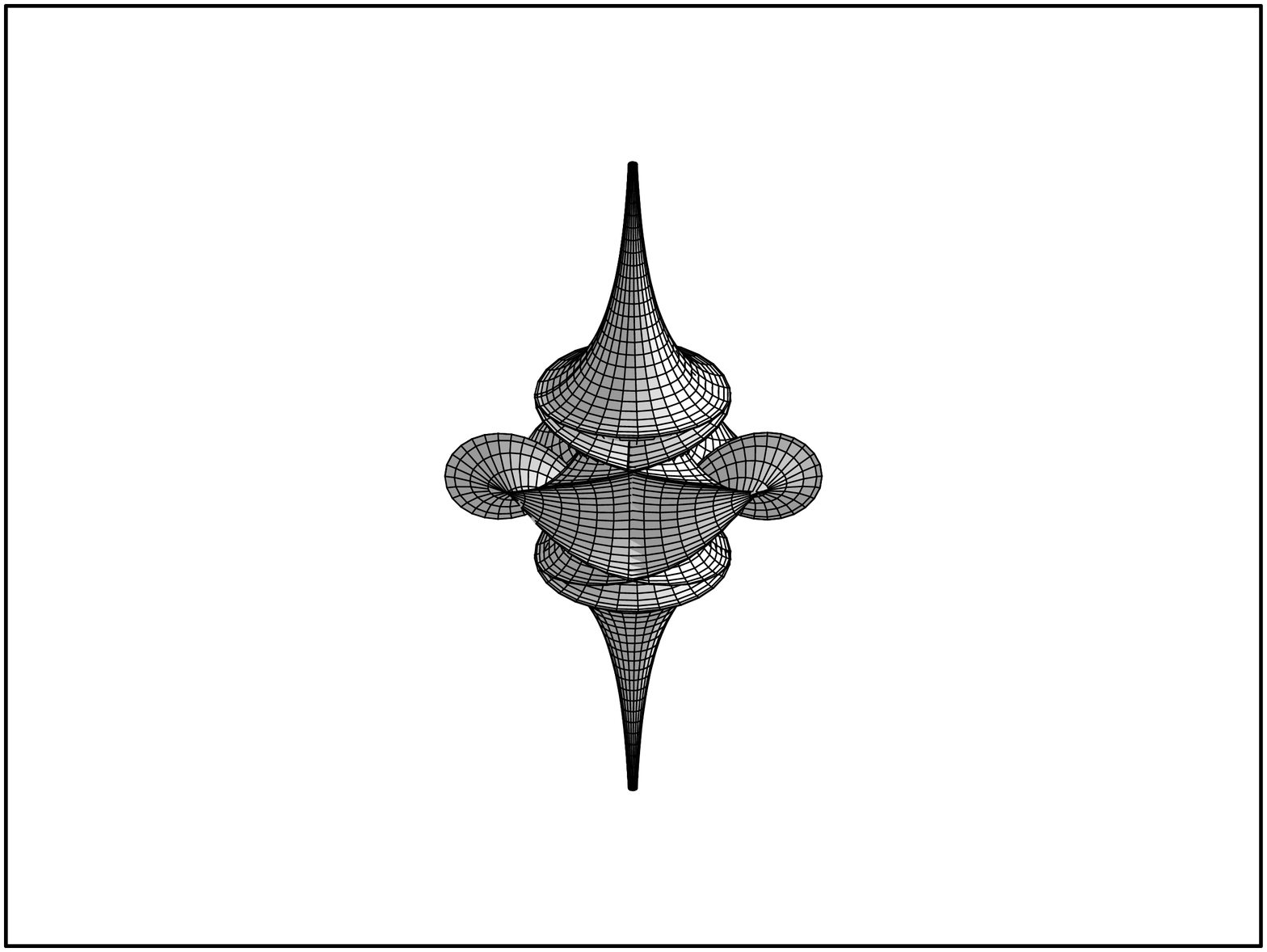}}
 \end{minipage}
 \begin{minipage}{0.49\textwidth}
 \centerline{\includegraphics[angle=270,trim= 110 190 110 190,width=\textwidth,
                              clip]{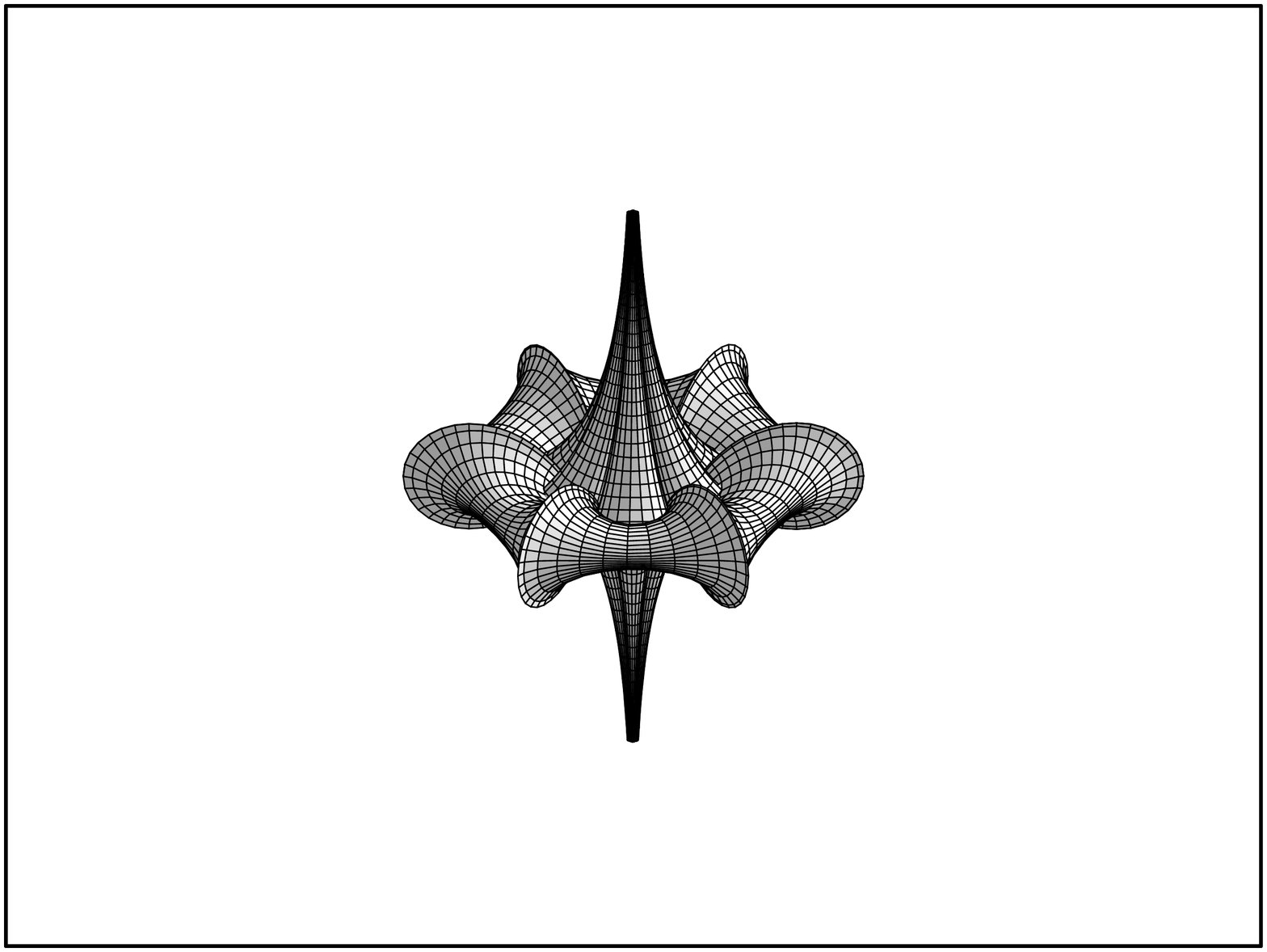}}
 \end{minipage}\\[2mm]
 \begin{minipage}{0.49\textwidth}
 \centerline{\includegraphics[angle=270,trim= 110 190 110 190,width=\textwidth,
                              clip]{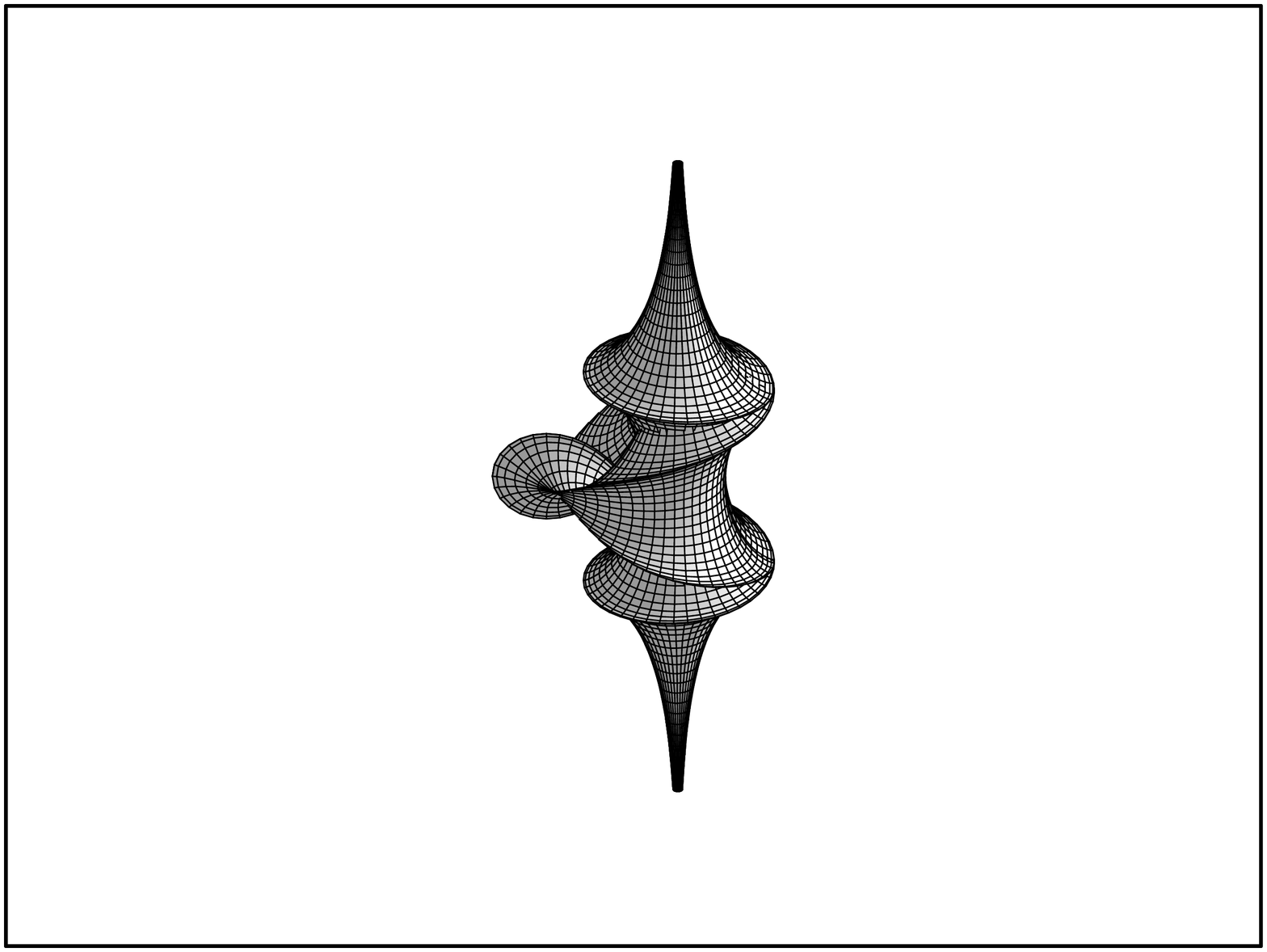}}
 \end{minipage}
 \begin{minipage}{0.49\textwidth}
 \centerline{\includegraphics[angle=270,trim= 110 190 110 190,width=\textwidth,
                              clip]{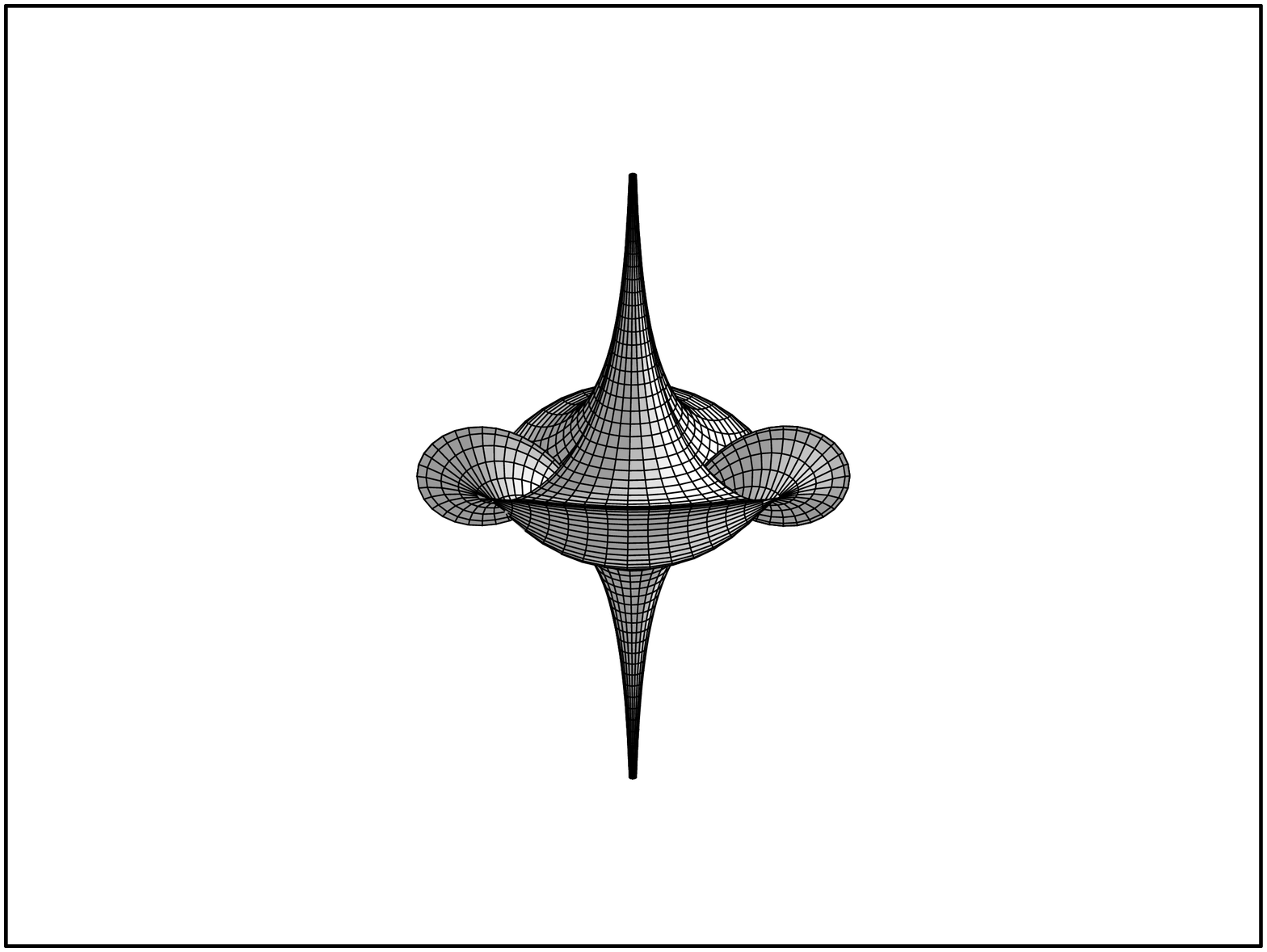}}
 \end{minipage}
 \caption{Breather pseudospherical surfaces: 
     $\frac{\mathsf{p}}{\mathsf{q}} 
          = \frac{1}{4},\frac{3}{4},\frac{1}{5},\frac{1}{2}$}
 \label{pseudobreathers}
\end{figure}

If $\lambda\neq 1$ then integration of the linear system (\ref{E81}) and 
specification of the constants of integration lead to the position vector
\bela{E83}
  \up{R}'_1 = \left(\bear{c}0\\ 0\\ x\ear\right) - \frac{2d}{c}
              \frac{\cosh cx}{c^2\sin^2dy + d^2\cosh^2cx}
              \left(\bear{c}-\sin dy\sin y - d\cos dy\cos y\\
                             \sin dy\cos y - d\cos dy\sin y\\ 
                             d\sinh cx\ear\right),
\end{equation}
where
\bela{84}
  \lambda = c^2, \quad c^2 + d^2 = 1.
\end{equation}
These pseudospherical surfaces are associated with the `stationary' {\em 
breather} solutions of the sine-Gordon equation (\ref{E78}) if the constants 
$c$ and $d$ are real. There exists a discrete rotational
symmetry if $d$ is rational, that is
\bela{E84}
  d = \frac{\mathsf{p}}{\mathsf{q}},\quad \mathsf{p},\mathsf{q}\in\mathbb{Z}.
\end{equation}
A variety of {\em breather pseudospherical surfaces} (Rogers \& Schief 2000)
corresponding to different choices of $\mathsf{p}$ and $\mathsf{q}$ is 
displayed in Figure \ref{pseudobreathers}.

It is interesting to note that the B\"acklund transformation for O surfaces
does not reduce to the {\em classical} B\"acklund transformation 
for pseudospherical surfaces. In fact, as discussed above, a single
application of the B\"acklund transformation for O~surfaces
to a straight line produces the pseudosphere or
breather pseudospherical surfaces. Here, it is required to solve a system of 
linear {\em non-autonomous} differential equations. By contrast, a single
application of the classical B\"acklund transformation involves the solution of
a {\em constant-coefficient} linear system and results in a 
one-parameter family of {\em Dini surfaces} including the 
pseudosphere~(Eisenhart 1960). 
A second application, which is purely algebraic in nature due to the existence
of an associated {\em permutability theorem} (Eisenhart 1960), then leads to
breather pseudospherical
surfaces if one assumes that the two B\"acklund parameters are complex 
conjugates (Rogers \& Schief 2000). 

\section{Concluding remark}

It is evident that any position matrix $\both{R}$
associated with O surfaces may be interpreted as a $3n$-dimensional vector
composed of the entries $R_{\kappa}^k$. Thus, $\both{R}$ defines a surface
$\Sigma$ in a $3n$-dimensional pseudo-Euclidean space endowed with a metric 
determined by the symmetric matrix $S$. If, for simplicity, we consider the
Euclidean case of a unit matrix $S$ with
\bela{E85}
  \down{H}^2 = 1,\quad\down{K}^2 = 1
\end{equation}
then the induced metric on $\Sigma$ is `flat', that is 
\bela{E86}
  {\rm I} =\, <d\both{R},d\both{R}>\, = dx^2 + dy^2.
\end{equation}
Canonical second fundamental forms are obtained by choosing an orthonormal
basis $\down{N}^{\alpha},\,\alpha = 3,\ldots n$ of the normal bundle associated
with the dual O surfaces $\Sigma_{\kappa}$ satisfying
\bela{E86a}
  \down{N}^{\alpha}_x = X^{\alpha}\down{H},\quad
  \down{N}^{\alpha}_y = Y^{\alpha}\down{K}
\end{equation}
with some functions $X^{\alpha}$ and $Y^{\alpha}$. Since
\bela{E87}
  \down{H}\cdot\down{N}^{\alpha} = 0,\quad \down{K}\cdot\down{N}^{\alpha} = 0,
  \quad\down{N}^{\alpha}\cdot\down{N}^{\beta} = \delta^{\alpha\beta},
\end{equation}
where $\delta^{\alpha\beta}$ denotes the usual Kronecker symbol, the quantities
\bela{E88}
 \bear{c}
  \both{N}_1^2 = \up{X}\down{K},\quad
  \both{N}_2^1 = \up{Y}\down{H},\quad
  \both{N}_3^1 = \up{N}\down{H},\quad
  \both{N}_3^2 = \up{N}\down{K}\as
  \both{N}_1^\alpha = \up{X}\down{N}^{\alpha},\quad
  \both{N}_2^\alpha = \up{Y}\down{N}^{\alpha},\quad
  \both{N}_3^\alpha = \up{N}\down{N}^{\alpha}
 \ear
\end{equation}
are readily shown to form an orthonormal basis of the $(3n-2)$--dimensional 
normal bundle attached to $\Sigma$. The corresponding second fundamental forms
\bela{E89}
  {\rm II}_k^\kappa = -<d\both{R},d\both{N}_k^\kappa>
\end{equation}
therefore become
\bela{E90}
 \bear{c}
  {\rm II}_1^2 = -q\,dx^2 + 2p\,dxdy - q\,dy^2,\quad
  {\rm II}_2^1 = -p\,dx^2 + 2q\,dxdy - p\,dy^2\as
  {\rm II}_3^1 = -H_{\circ}dx^2,\quad
  {\rm II}_3^2 = -K_{\circ}dy^2\as
  {\rm II}_1^\alpha = -X^\alpha dx^2,\quad
  {\rm II}_2^\alpha = -Y^\alpha dy^2,\quad
  {\rm II}_3^\alpha = 0.
 \ear
\end{equation}
The geometry of the surfaces $\Sigma\subset\mathbb{R}^{3n}$ defined via
Combescure-related O~surfaces is currently being investigated.

\medskip
\noindent
{\bf Acknowledgement.} One of the authors (B.G.K.) is grateful to the School
of Mathematics, UNSW for the very kind hospitality.

\end{document}